# Atomic level understanding of site-specific interactions in Polyaniline/TiO$_2$ composite

Satyananda Chabungbam,[a,b] G. C. Loh,[b,c] Munima B Sahariah,[a]* Arup Ratan Pal,[a] Ravindra Pandey [b]*

[a] Institute of Advanced Study in Science and Technology, Guwahati-781035, INDIA

[b] Physics Department, Michigan Technological University, Houghton-49931, MI, USA

[c] Institute of High Performance Computing, Singapore-138632

(October 15, 2015)

*E-mail: munima@iasst.gov.in, pandey@mtu.edu

**Abstract**

The results of spin-polarized density functional theory calculations find that band gap engineering can be achieved by site-specific interactions in a composite consisting of polyaniline and $TiO_2$ nanoparticles. Interactions in the composite matrix are found to be mediated by Ti atoms inducing dependency of location of the conduction band minimum on the polyaniline site which is being probed by $TiO_2$. This dependency is due to subtle changes in the nature of valance or conduction states near Fermi level introduced by the interacting matrix sites. The results therefore suggest that optimization of the synthesis parameters at atomic level can be an effective way to improve performance of a photovoltaic device based on PAni- $TiO_2$ composite.

# 1.0 Introduction

Polymer composites consisting of organic and inorganic materials generally exhibit unusual properties which differ from properties exhibited by their component materials. These hybrid composites are promising candidates for applications in energy and health related areas. For example, the hybrid composite of polyaniline and $TiO_2$ has been suggested for applications in diverse areas including electrochemical capacitors, heterostructure devices, microbial fuel cells, and gas sensing devices [1-4]. It is to be noted that polyaniline (PAni) is one of the oldest organic conducting polymers known for its stability, ease of processing, doping/dedoping flexibility and low cost of manufacturing [5]. $TiO_2$ is one of the technologically important oxide semiconductors with strong potential applications in a variety of devices [6-9].

In recent years, photosensitive devices based on organic semiconductors have been pursued as viable alternatives to silicon photovoltaics. Growing research on organic photovoltaics (OPV) has shown that they might be able to address future energy supply issues due to their low production cost, flexibility and roll-to-roll printing methods [10, 11]. Various prototypes of hybrid photodetectors have since been fabricated and tested. Photodetectors based on PAni and $TiO_2$ hybrid composites, both under photoconductive and photovoltaic modes have shown excellent results including high photosensitivity, fast response speed and good environmental stability [4, 12-14].

An ideal scenario for fabrication of a photovoltaic device using PAni and $TiO_2$ would be $TiO_2$ nanoparticles well-dispersed in the polymer matrix to form percolated networks, which could then improve the device performance manifold. In a typical device architecture with the photoactive layer as the PAni-$TiO_2$ composite, PAni serves as the donor and $TiO_2$ serves as the acceptor material. The light generation and collection process can then involve the following steps. At first, light absorption by the active material (donor/acceptor composite) takes place followed by generation of excitons. Next, charge separation occurs at the interface between donor and acceptor due to high electron affinity of the acceptor material. Subsequently, transport of charges through the bulk material takes place, and finally the charges are collected at the respective electrodes resulting in a flow of current under the external field. Proper band alignment, interfacial properties between the donor/acceptor materials and tuned bandgap are three predominant factors on which the performance of a photovoltaic device relies. Although different hybrid materials are synthesized and utilized for fabrication of the stable photovoltaic devices, question of bandgap tuning of individual material is still challenging. This is due to the fact that material scientists are now seeking more facile and efficient techniques to shift the optical absorption of semiconducting materials to the visible-infrared regions of the electromagnetic spectrum. In particular, in spite of having greater stability, the band gap of PAni-TiO$_2$ hybrid composite is difficult to tune in the visible-infrared range [13, 14], and hence the utilization of PAni-TiO$_2$ composite in efficient photovoltaic applications is somewhat limited. Appropriate bandgap tuning may also facilitate the utilization of this hybrid composite as electrochromic materials and in visible light induced photocatalysis [15-18]. In this paper, we focus on tailoring the magnitude of the band gap via site-specific interactions at nanoscale in PAni-TiO$_2$ composite. In particular, electronic structure calculations will be performed on the fully reduced state of PAni and sub-nm sized nanoparticles in the form of a (TiO$_2$)$_3$ cluster to provide atomic-level understanding of the site-specific interactions in the hybrid composite.

**2.0 Computational Method**

Calculations based on density functional theory were performed using the Vienna Ab-initio Simulation Package (VASP) with Projector Augmented Wave (PAW) method. [19-22] The Perdew-Burke-Ernzerhof (PBE) [23] functional form under the generalized gradient approximation (GGA) was used. In calculations, the polyaniline chain was extended in the x-direction, and the dimensions of the supercell were (21.3x20x20) Å$^3$. In the periodic supercell, a vacuum distance normal to the plane was set to be larger than 20 Å to eliminate interaction between the image replicas. The plane wave energy cut-off was fixed at 500 eV. The reciprocal space was sampled by a grid of (8x1x1) k-points in the Brillouin zone. The energy and force convergence criteria were fixed at 10$^{-6}$ eV and 0.001 eV/A, respectively. Bader's charge [24-26] analysis was performed considering charge contributions from both valence and core electrons. Analyses of the bonding and antibonding states via the crystal orbital Hamilton population (COHP) were performed using the LOBSTER package [27-29].

PAni has three oxidation states: (i) Leucoemeraldine base (LB) in the fully reduced state, (ii) Emeraldine base (EB) in the partially oxidized state, and (iii) Pernigraniline base (PB) in the fully oxidized state [30]. The calculated results find the LB form of PAni to be the most stable in agreement with the previous theoretical results [30, 31], thus showing reliability of our approach based on the GGA-DFT level of theory. All the atomic structures and charge densities are plotted with VESTA 3 visualization package [32].

### 3.0 Results and Discussion

#### 3.1 Polyaniline (PAni)

Figure 1 shows the equilibrium configuration of LB-PAni with the calculated bond lengths of $R_{C1-C2}$= 1.41, $R_{C2-C3}$= 1.39, $R_{C3-C4}$= 1.41, $R_{C4-C5}$= 1.41, $R_{C5-C6}$= 1.39, $R_{C6-C1}$= 1.41, $R_{C1-N}$= 1.40, and $R_{C4-C2}$= 1.40 Å which are in excellent agreement with previous DFT results [31].

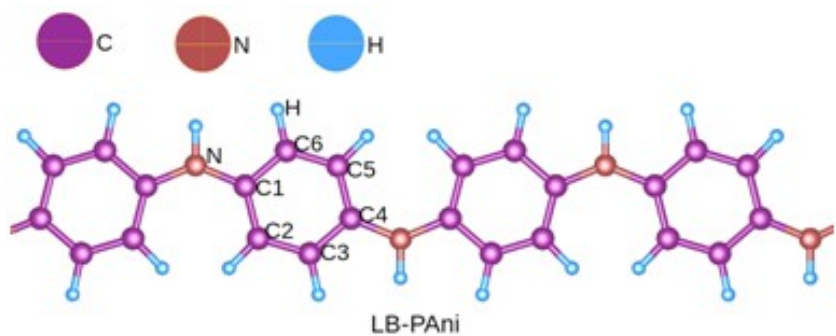

Figure 1: The equilibrium configuration of LB-PAni.

The calculated band gap of LB-PAni is 1.8 eV whereas the corresponding experimental value is 3.6-3.8 eV [33]. This underestimation of the band gap is a well-known deficiency in DFT-GGA calculations, though the experimental samples are also affected by imperfections in synthesis conditions. For LB-PAni of a finite chain-length, previous DFT calculations have predicted its gap to be 4-5 eV [34, 35]. It should be noted that the energy difference between planar and nonplanar LB-conformations is found to be rather small of about 0.015 eV/atom. Therefore, the planar structure of LB-PAni is considered for further calculations allowing us to utilize our modest computational resources for calculations without modifying physics and chemistry of the system.

## 3.2 Sub-nm sized Nanoparticles: $(TiO_2)_3$

Total energy calculations were performed on planar and non-planar conformers of the sub-nm sized nanoparticles represented by a $(TiO_2)_3$ cluster. Figure 2 shows the ground state of $(TiO_2)_3$ consisting of a cage-like configuration with $R_{(Ti-O)}$ ranging from 1.68 to 2.08 Å. $R_{(Ti-Ti)}$ is calculated to be 2.82-2.85 Å in agreement with the previous study [36]. Note that the planar $(TiO_2)_3$ conformer is 2.3 eV higher in energy than the non-planar cage-like $(TiO_2)_3$.

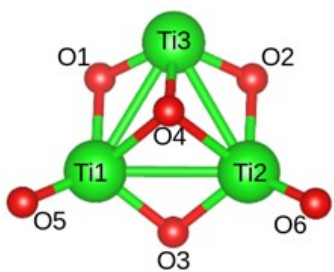

Figure 2: The ground state configuration of (TiO2)3.

## 3.3. Polyaniline -$TiO_2$ Composite

Several sites of PAni are considered to determine the most energetically preferential site for $(TiO2)_3$ in the matrix. They are classified as either top, bridge or hollow sites, such as $Top_{C1}$, $Top_{C2}$, $Top_N$, $Top_H$, Bridge1 (bridge between C1 and C2), Bridge2 (bridge between C2 and N) and Hollow1 (hollow site at the center of hexagon) as shown in Figure 3.

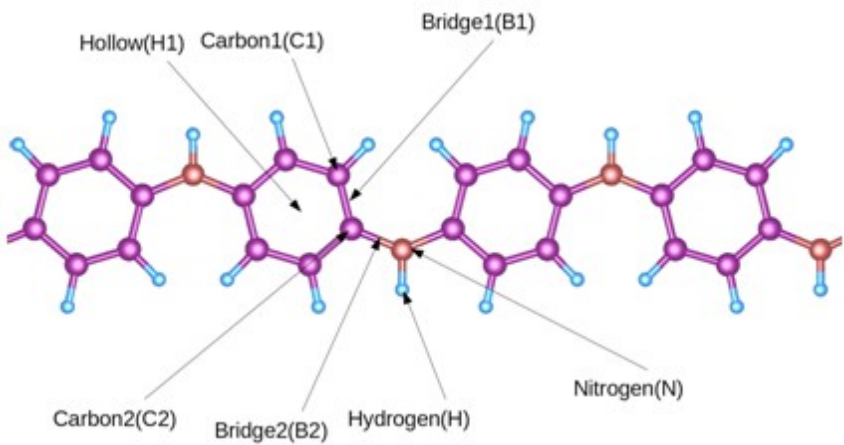

Figure 3. Schematic of PAni sites considered for interaction with (TiO2)3. (C-Violet, N-Brown, H-Blue)

The energy surfaces representing the interaction of (TiO2)₃ with PAni are obtained by varying the vertical height of (TiO2)₃ with respect to the plane at each interacting PAni site. Two orientations of (TiO2)₃ with respect to the PAni are considered: (i) the vertex Ti atom nearer the PAni plane (Figure 4), and (ii) the central O atom nearer the PAni plane (Supplementary Information, Figure S1). In the former, the central axis of the TiO$_2$ passing through the vertex Ti atom makes an angle of 64º relative to the PAni plane.

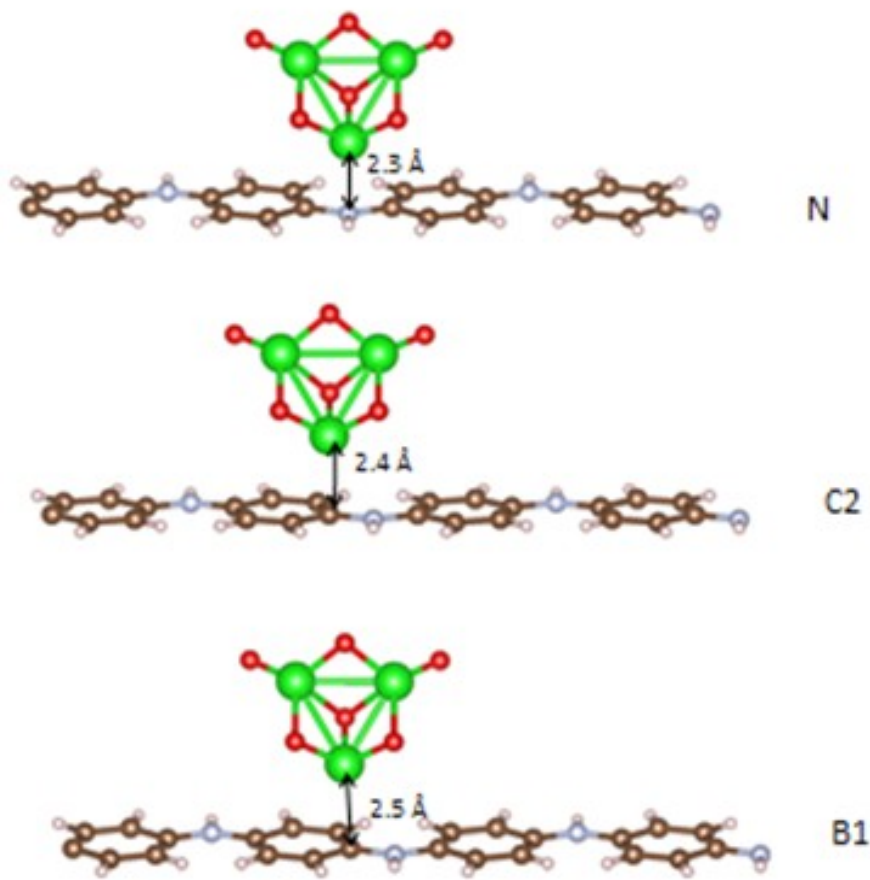

Figure 4. Schematic of the equilibrium configurations of (TiO$_2$)$_3$ interacting with PAni at Bridge (B1), Top$_{C2}$ and Top$_N$ sites. (Ti: large green circle, O: small red circle).

In an equilibrium configuration, the binding energy is defined in terms of the constituents of the hybrid system. For Ti-orientation, Top$_{H1}$ site of PAni is least preferred with the binding energy of 0.1 eV. The remaining interacting sites (i.e. Top$_{C1}$, Top$_{C2}$, Top$_N$, Bridge1, Bridge2 and Hollow1 sites) of PAni are almost equally preferred with binding energies of 0.15-0.17 eV.

Interestingly, O-orientation of $(TiO_2)_3$ approaching PAni is not predicted to be the energetically preferential with the binding energy values of 0.03-0.5 eV (Table S1). The calculated results therefore find the interaction between $(TiO_2)_3$ and PAni to be mediated by Ti atoms in the matrix. It should be noted that the results from spin polarized calculations find the hybrid system to be non-magnetic.

In order to understand the predicted affinity of Ti atoms with the PAni matrix, we have calculated the charge density and electron localization functions associated with Bridge1, $Top_{C2}$ and $Top_N$ sites exhibiting similar degree of affinity with $(TiO_2)_3$. Figure 5 shows the electron localization function (ELF) from two different planes; one plane parallel to the plane of PAni and the other perpendicular to PAni plane intersecting the midpoint of the cluster. An ELF of magnitude $\eta = 0.65$ is found to be suitable to reproduce the localized electron pairs for all sites in the hybrid system. The localization domains near the center point of the C-C bonds suggest the highly covalent nature of the bonds. Such domains are absent for the C-N, N-H, and C-H bonds. Additionally, Bader charge analysis [24-26] finds that there is an overall increase of $\approx 0.20|e|$ for $(TiO_2)_3$ indicating charge transfer from the donor (i.e. PAni matrix) to the acceptor (i.e. $(TiO_2)_3$) thus increasing ionicity between Ti and O atoms in the hybrid system ((Supplementary Information, Table S2).

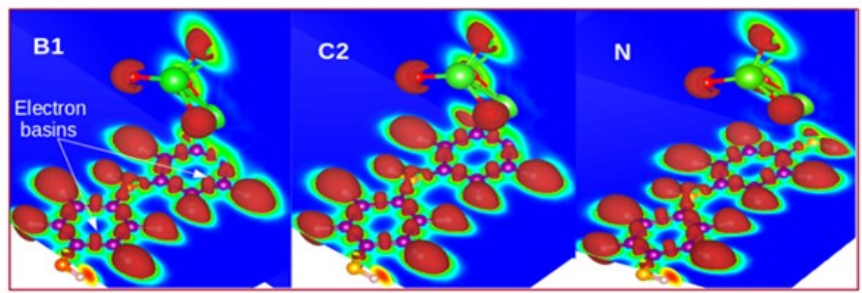

Figure 5: Electron localization function of the PAni-$TiO_2$ composite at Bridge1, $Top_{C2}$ and $Top_N$ sites. Localized electron pairs known as electron basins are formed in between C-atoms whereas extended basins are formed between C-H and N-H bonds.

The effect of small but noticeable change in the nature of bonds within the hybrid system is reflected in total density of states given in Figure 6. For all cases, semiconducting nature of the hybrid system is predicted. The valence band maximum (VBM) region mainly consists of states associated with atoms of PAni matrix. This is not the case with the conduction band minimum (CBM) which is dominated by states associated with Ti atoms. And the location of CBM is found to depend on the $TiO_2$ binding site of the matrix yielding a band gap of 0.97 eV, 2.38 eV and 1.65 eV for Bridge1, $Top_{C2}$ and $Top_N$ configurations, respectively. It is important to point out here that recently site dependent doping method has been used to tune a variety of material properties like thermoelectric parameters, spin manipulation, charge transport, etc [37-39].

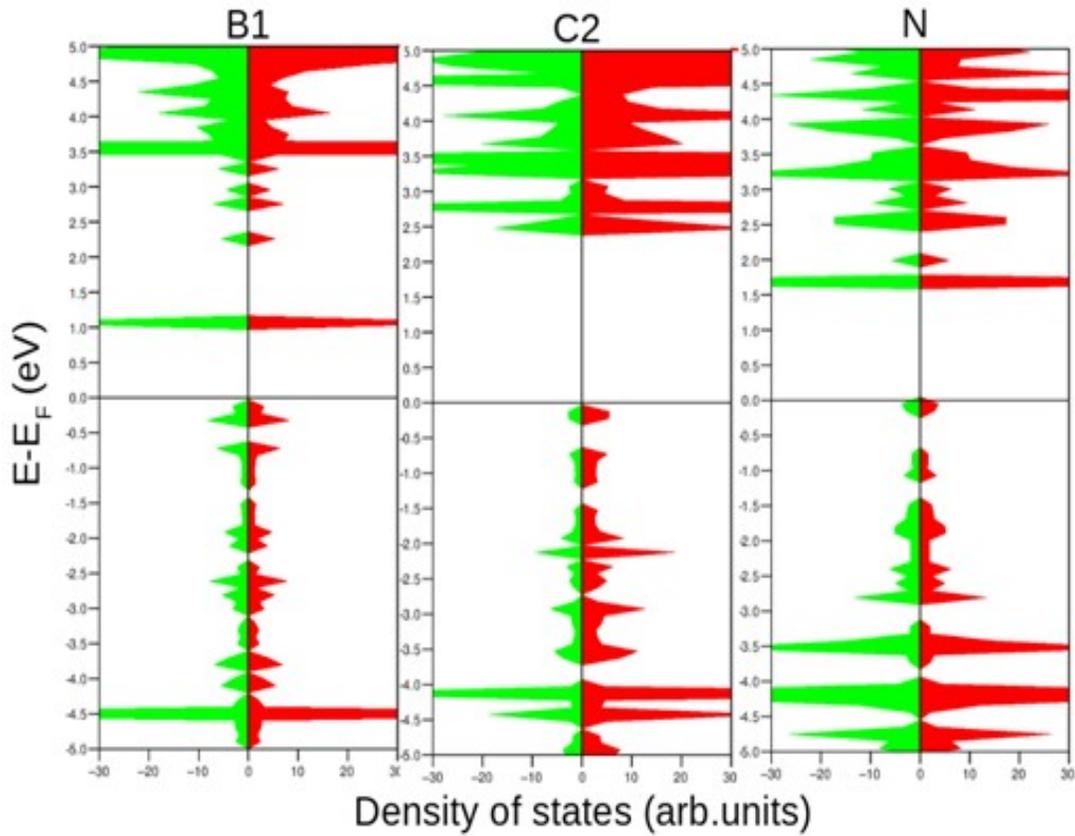

Figure 6: Spin-polarized DOS of PAni-$TiO_2$ associated with Bridge1, $Top_{C2}$ and $Top_N$ configurations. The red region represents the spin-up states while the green region represents the spin-down states. Zero is taken to be the Fermi level.

To see the interactions in the lower conduction bands, we calculate the partial charge densities at different energy ranges. Figures 7(a) and 7(b) show the partial charge density distributions at the range 1-1.3 eV and 2-3 eV at Bridge1-site. Figures 7(c) and 7(d) show the partial charge density distribution at the range 2-3 eV at $Top_{C2}$ and $Top_N$ sites. From the figures, it is clear that the process of charge transfer from PAni-matrix to $TiO_2$ cluster in the specified energy windows are different for different doping sites. This ultimately give rise to dependency of the CBM on the binding site of $TiO_2$. In order to gain additional insights, we further examine the nature of bonding between PAni and $(TiO_2)_3$ using the crystal orbital Hamilton population (COHP) method. Figure 8 shows the results of COHP analysis for the hybrid system where the positive population represents the bonding states while the negative population represents the antibonding states in the system. At the bridge site of the hybrid system, Ti-atom has two nearest neighbors C atoms.

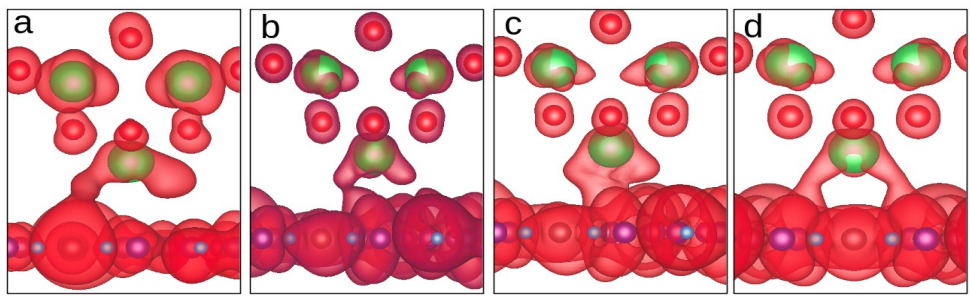

Figure 7 :   Conduction state charge densities in the energy windows (a) (1-1.3) eV at Bridge1-site;  (b) (2-3) eV at Bridge1-site; (c) (2-3) eV at $Top_{C2}$-site; (d) (2-3) eV at $Top_N$-site

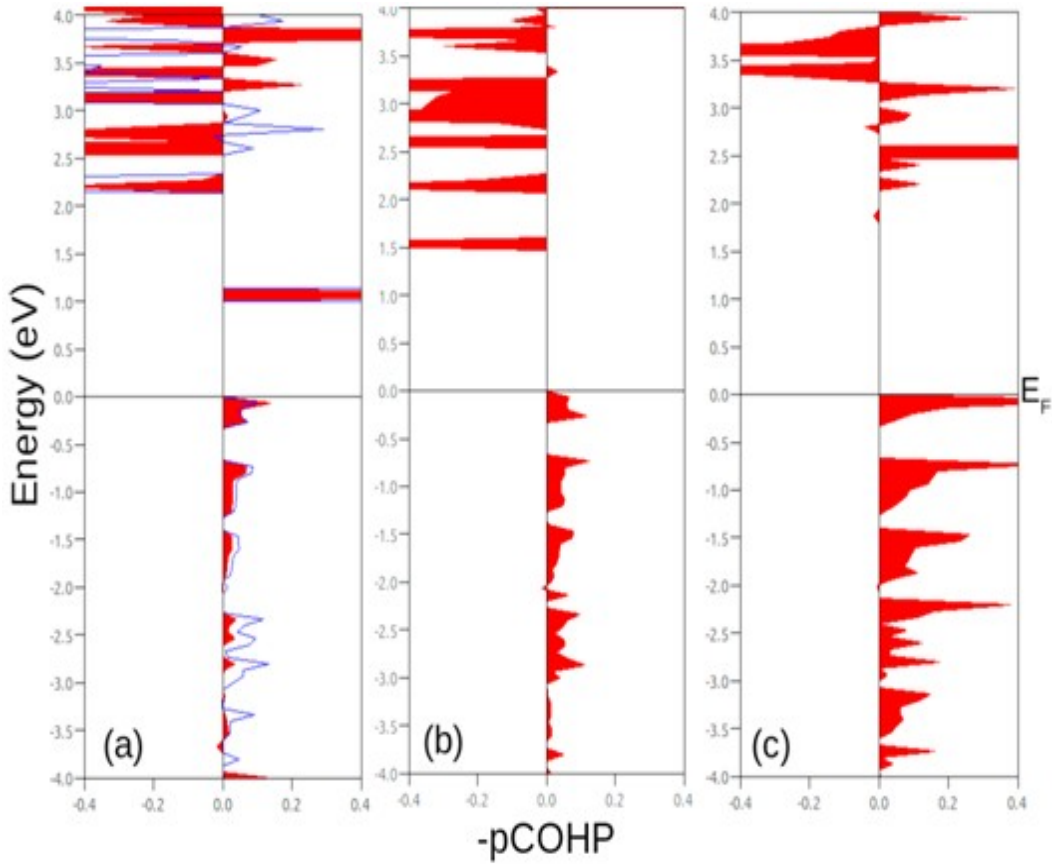

Figure 8: pCOHP plots of the hybrid system: (a) Bridge site, Ti-C1(red region)/Ti-C2(blue lines) (b) $Top_{C2}$-site, Ti-C2 (c) $Top_N$-site, Ti-N. The positive population represents the bonding states while the negative population represents the antibonding states. Zero is taken to be the Fermi level.

Contributions from interactions between Ti-C1 and Ti-C2 atoms are shown by the red and blue regions, respectively. These contributions are similar in the lower bonding regions, but are slightly different for the higher antibonding states. The bonding states even cross the Fermi level with a peak around 1 eV. This is not the case with the $Top_{C2}$-site, where the bonding states do not cross the Fermi level and the higher energy states are the antibonding states arising due to contributions from interacting neighboring atoms, Ti and C in the hybrid system. For the $Top_N$-site, there appears a mix of the bonding-antibonding states at the higher energies. Thus, the COHP analysis clearly shows a subtle dissimilarity in nature of bonding at the PAni sites interacting with $(TiO_2)_3$ which may lead to site-dependency of the position of CBM in the hybrid system.

## 4.0 Summary

First principles calculations based on density functional theory are performed to provide atomic level understanding of site-specific interactions in a polymer composite consisting of polyaniline and sub-nm sized TiO$_2$ particles. We find that the bonding in such a polymer composite will be mediated by Ti atoms, and subtle changes in the nature of bonds between PAni and (TiO$_2$)$_3$ can lead to variation in the band gap of the composite system. The calculated results, thus, give a qualitative support to the fact that the bandgap of an organic-inorganic hybrid material can be finely tuned by site-specific interactions in the polymer matrix.

## AUTHOR INFORMATION

**Corresponding Authors**
*Email: munima@iasst.gov.in, pandey@mtu.edu

**Author Contributions**
The manuscript was written through contributions of all authors.

**Notes**
The authors declare no competing financial interest.

## ACKNOWLEDGMENT
Helpful Discussion with Dr. D. S. Patil and Munish Sharma are acknowledged. All the calculations were carried out at Michigan Tech's computer cluster RAMA. CS acknowledges Department of Physics, Michigan Technological University, USA for providing support to visit MTU.

**REFERENCES**

1. Huyen, D. N. et al., Effect of TiO2 on the gas sensing features of TiO2/PANi nanocomposites. *Sensors*, **2011**, 11(2): p. 1924-1931.

2. Li, X. et al., A novel electrode material based on a highly homogeneous polyaniline/titanium oxide hybrid for high-rate electrochemical capacitors. *J. Mater. Chem.*, **2010**, 20(47): p. 10598-10601.

3. Qiao, Y. et al., Nanostructured polyaniline/ titanium dioxide composite anode for microbial fuel cells. *ACS Nano*, **2008**, 2(1): p. 113-119.

4. Hussain, A. A.; Pal A. R.; and Patil D. S., An efficient fast response and high-gain solar-blind flexible ultraviolet photodetector employing hybrid geometry. *Appl. Phys. Lett.*, **2014**, 104(19): p. 193301.

5. Okamoto, Y. and Brenner, W., *Organic semiconductors* **1964**: Reinhold Pub. Corp.

6. Heller, A. Hydrogen-evolving solar cells. *Science*, **1984**, 223(4641): p. 1141-1148.

7. Gratzel, M. Photoelectrochemical cells. *Nature*, **2001**, 414(6861): p. 338-344.

8. Bavykin, D. V.; Friedrich, J. M.; and Walsh, F. C. Protonated titanates and TiO2 nanostructured materials: synthesis, properties, and applications. *Adv. Mater.*, **2006**, 18(21): p. 2807-2824.

9. Gratzel, M. Dye-sensitized solar cells. *J. Photochem. Photobiol. C: Photochemistry Reviews*, **2003**, 4(2): p. 145-153.

10. Dennler, G.; Scharber, M. C.; and Brabec, C. J.; Polymer-fullerene bulk-heterojunction solar cells. *Adv. Mater.*, **2009**, 21(13): p. 1323-1338.

11. Krebs, F. C.; Gevorgyan, S. A.; and Alstrup, J. A roll-to-roll process to flexible polymer solar cells: model studies, manufacture and operational stability studies. *J. Mater. Chem.*, **2009**, 19(30): p. 5442-5451.

12. Hussain, A. A. et al. Comparative study of nanocomposites prepared by pulsed and dc sputtering combined with plasma polymerization suitable for photovoltaic device applications. *Mater. Chem. Phys*., **2014**, 148(3): p. 540-547.

13. Hussain, A. A.; Pal, A. R. and Patil, D. S. High photosensitivity with enhanced photoelectrical contribution in hybrid nanocomposite flexible UV photodetector. *Organic Electronics*, **2014**, 15(9): p. 2107-2115.

14. Baro, M.; Hussain, A. A.; and Pal, A. R. Enhanced light sensing performance of a hybrid device developed using as-grown vertically aligned multiwalled carbon nanotubes on TCO substrates. *RSC Advances*, **2014**, 4(87): p. 46970-46975.

15. Cai, G. et al., Multicolor electrochromic film based on TiO2@ polyaniline core/shell nanorod array. *J. Phys. Chem. C*, **2013**, 117(31): p. 15967-15975.

16. Xiong, S. et al., Covalently bonded polyaniline− TiO2 hybrids: A facile approach to highly stable anodic electrochromic materials with low oxidation potentials. *Chem. Mater*., **2010**, 22(1): p. 255-260.

17. Lin, Y. et al., Highly efficient photocatalytic degradation of organic pollutants by PANI-modified TiO2 composite. *J. Phys. Chem. C*, **2012**, 116(9): p. 5764-5772.

18. Zhang, H. et al. Dramatic visible photocatalytic degradation performances due to synergetic effect of TiO2 with PANI. *Environ. Sci. Technol.*, **2008**, 42(10): p. 3803-3807.

19. Kresse, G. and Furthmuller, J. Efficiency of ab-initio total energy calculations for metals and semiconductors using a plane-wave basis set. *Comput. Mater. Sci.*, **1996**, 6(1): p. 15-50.

20. Kresse, G. and Furthmuller J. Efficient iterative schemes for ab initio total-energy calculations using a plane-wave basis set. *Phys. Rev. B*, **1996**, 54(16): p. 11169.

21. Kresse, G. and Hafner, J. Ab initio molecular dynamics for liquid metals. *Phys. Rev. B*, **1993**, 47(1): p. 558.

22. Vanderbilt, D. Soft self-consistent pseudopotentials in a generalized eigenvalue formalism. *Phys. Rev. B*, **1990**, 41(11): p. 7892.

23. Perdew, J. P.; Burke, K.; and Ernzerhof, M. Generalized gradient approximation made simple. *Phys. Rev. Lett.*, **1996**, 77(18): p. 3865.

24. Tang, W.; Sanville, E.; and Henkelman, G. A grid-based Bader analysis algorithm without lattice bias. *J. Phys.: Condens. Matter*, **2009**, 21(8): p. 084204.

25. Sanville, E. et al., Improved grid-based algorithm for Bader charge allocation. *J. Comp. Chem.*, **2007**, 28(5): p. 899-908.

26. Henkelman, G.; Arnaldsson, A.; and Jonsson H., A fast and robust algorithm for Bader decomposition of charge density. *Comput. Mater. Sci.*, **2006**, 36(3): p. 354-360.

27. Dronskowski, R. and Bloechl, P. E. Crystal orbital Hamilton populations (COHP): energy-resolved visualization of chemical bonding in solids based on density-functional calculations. *J. Phys. Chem.*, **1993**, 97(33): p. 8617-8624.

28. Deringer, V. L.; Tchougreeff, A. L.; and Dronskowski, R. Crystal orbital Hamilton population (COHP) analysis as projected from plane-wave basis sets. *J. Phys. Chem. A*, **2011**, 115(21): p. 5461-5466.

29. Maintz, S. et al. Analytic projection from plane-wave and PAW wavefunctions and application to chemical-bonding analysis in solids. *J. Comput. Chem.*, **2013**, 34(29): p. 2557-2567.

30. Gustavo, M. and Nascimento, D. *Spectroscopy of Polyaniline Nanofibers* **2010**: INTECH Open Access Publisher.

31. Zheng, G., et al. Lattice dynamics of polyaniline and poly (p-pyridyl vinyline): First-principles determination. *Phys. Rev. B*, **2006**, 74(16): p. 165210.

32. Momma, K. and Izumi, F. VESTA 3 for three-dimensional visualization of crystal, volumetric and morphology data. *J. Appl. Crystallogr.*, **2011,** 44: p. 272-1276.

33. Mullen, K. and Wegner G. *Electronic Materials: The Oligomer Approach*. Wiley-VCH: Weinheim **1998**

34. Mishra, A. K. and Tandon P., A comparative ab initio and DFT study of polyaniline leucoemeraldine base and its oligomers. *J. Phys. Chem. B*, **2009**, 113(44): p. 14629-14639.

35. Aleman, C., et al., On the molecular properties of polyaniline: A comprehensive theoretical study. *Polymer*, **2008**, 49(23): p. 5169-5176.

36. Cakir, D. and Gulseren, O. Ab initio study of neutral (TiO2) n clusters and their interactions with water and transition metal atoms. *J. Phys.: Condens. Matter*, **2012**, 24(30): p. 305301.

37. Abutaha A. I. et al. Doping site dependent thermoelectric properties of epitaxial strontium titanate thin films. *J. Mater. Chem. C*, **2014**, 2: p. 9712

38. Pi S. T. et al. Site-dependent doping effects on quantum transport in zigzag graphene nanoribbons. *Carbon,* **2015,** 94: p. 196–201

39. Cornelius K. et al. Site- and orbital-dependent charge donation and spin manipulation in electron-doped metal phthalocyanines. *Nature Mat.*, **2013**, 12: p. 337-343.

Graphical TOC Entry

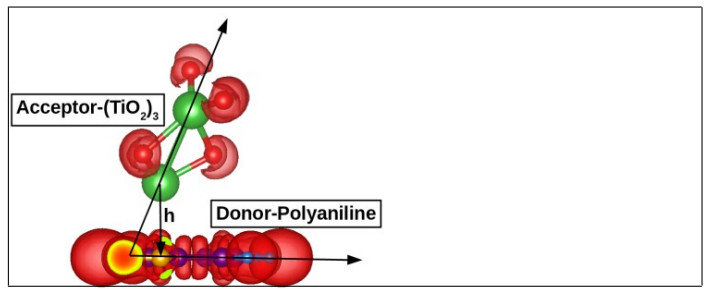